\documentclass[aps,prb,twocolumn,showpacs,superscriptaddress]{revtex4}
\usepackage{amsmath, amsfonts,amssymb,bbm}
    \usepackage{graphicx}
    \usepackage{hyperref}
    \usepackage{color, soul}
    \usepackage{tikz}
\DeclareRobustCommand*{\citen}[1]{%
  \begingroup
    \romannumeral-`\x % remove space at the beginning of \setcitestyle
    \setcitestyle{numbers}%
    \cite{#1}%
  \endgroup
}
    
\begin{document}
\title{Conduction in quasi-periodic and quasi-random lattices: \\Fibonacci, Riemann, and Anderson models} 
 \author{V.~K.~Varma} \affiliation{The \textit{Abdus Salam} International Centre for Theoretical Physics, Strada Costiera 11, 34151 Trieste, Italy}
 \author{S.~Pilati} \affiliation{The \textit{Abdus Salam} International Centre for Theoretical Physics, Strada Costiera 11, 34151 Trieste, Italy}
 \author{V.~E.~Kravtsov}
 \affiliation{The \textit{Abdus Salam} International Centre for Theoretical Physics, Strada Costiera 11, 34151 Trieste, Italy}
 \affiliation{L. D. Landau Institute for Theoretical Physics, Chernogolovka, Russia} 
\date{\today}

\vspace*{-1cm}
\begin{abstract}
We study the ground state conduction properties of noninteracting electrons in aperiodic but non-random one-dimensional models with chiral symmetry, 
and make comparisons against Anderson models with non-deterministic disorder. 
The first model we consider is the Fibonacci lattice, which is a paradigmatic model of quasicrystals; the 
second is the Riemann lattice, which we define inspired by Dyson's proposal on the possible connection between the Riemann hypothesis and a suitably defined quasicrystal.
Our analysis is based on Kohn's many-particle localization tensor defined within the modern theory of the insulating state.
In the Fibonacci quasicrystal, where all single-particle eigenstates are critical (i.e., intermediate between ergodic and localized), the noninteracting electron gas is found to be a conductor at most electron 
densities, including the half-filled case; however, at various specific fillings $\rho$, including the values $\rho = 1/g^n$, where $g$ is the golden ratio and $n$ is any integer, 
the gas turns into an insulator due to spectral gaps.
Metallic behaviour is found at half-filling in the Riemann lattice as well; however, in contrast to the Fibonacci 
quasicrystal, the Riemann lattice is generically an insulator due to single-particle eigenstate localization, likely at \textit{all} other fillings. 
Its behaviour turns out to be alike that of the off-diagonal Anderson model, albeit with different system-size scaling of the band-centre anomalies.
The advantages of analysing the Kohn's localization tensor instead of other measures of localization familiar from the theory of Anderson insulators 
(such as the participation ratio or the Lyapunov exponent) are highlighted.
\end{abstract}
\maketitle
Explaining and predicting the conduction properties of quantum systems which are neither periodic nor purely random is a challenging problem.
The most relevant example of such systems are the quasicrystals. These are materials that fall outside the conventional definition of crystals which $-$ in its superseded version $-$ subsumed only 
periodic real space structures. Quasicrystals are quasiperiodic in the sense that, while the translation symmetry is not preserved, they display well-defined diffraction patterns 
\cite{Shechtman, Steinhardt}, similar to periodic structures. 
Formally, the Fourier transform of their density distribution must contain, albeit perhaps dense, \textit{at least} another point spectrum:
\begin{equation}
 \label{eq: RiemannFourier}
 \mathcal{F}\left \{ \sum_{\gamma_n \in X}\delta_D(\gamma - \gamma_n) \right \} = \sum_{k_m \in X^{*}} F_{m} \delta_D(k - k_{m}),
\end{equation}
where $\delta_{\textrm{D}}$ is the Dirac delta function, and $\gamma_n$ and $k_m$ are in the countable subsets $X$ and $X^{*}$, in real space and in Fourier space, respectively~\cite{Senechal, Hof,Notediffraction}. 
Note that the right-hand side of Eq. \eqref{eq: RiemannFourier} might also include a continuous part~\cite{AubryLuck, Senechal}.

The first material discovered displaying such exotic diffraction properties was a metallic alloy of Al-Mn \cite{Shechtman, Steinhardt}, followed by alloys such as GaAs \cite{MerlinExpt} and $\textrm{Si-Ge}_x\textrm{Si}_{1-x}$ \cite{Houghton}. 
The conduction properties of most quasicrystals are unconventional and evasive: while they are (bad) metals at low temperatures \cite{Dubois}, their resistivity can decrease with increasing temperature, 
contrary to what is observed in most metals \cite{Martin}.
In the present day, quasicrystals may also be realized in cold-atom set-ups \cite{Roati, Bloch, Singh} and photonic waveguide experiments \cite{Lahini,Mher}, providing us with a new controllable 
experimental set-up to study the physics of these aperiodic structures.

Most theoretical studies on quasicrystals have focussed on the spectral properties and on the nature of the single-particle wave-functions, considering in particular the archetypal example of the 
Fibonacci lattice \cite{Kadanoff, Siggia, Fujiwara}. 
It has been established that the spectrum of this system is a Cantor set of zero measure, that it is purely singular continuous, and that it hosts  a dense set of gaps. 
Its single-particle eigenstates are known to be critical, being intermediate between the extended states, characteristic of clean periodic systems, and the exponentially localized states 
characteristic of Anderson localized systems with (strong) non-deterministic disorder.
However, the conduction properties of electrons in quasicrystalline systems such as the Fibonacci lattice are still very poorly understood \cite{Dubois, Rotenberg}. In fact, 
these systems represent a severe theoretical challenge: while the band structure theory based on periodic Bloch function has to be abandoned, 
the theory of the Anderson transition \cite{Anderson, Gang4}, which was developed for non-deterministic disordered systems with gapless spectra and localized single-particle eigenfunctions, 
is not directly applicable either.
%%%%%%%%%%%%%%%%%%%%%%

In this article we investigate the metal to insulator transition in one-dimensional quasi-disordered (i.e. intermediate between periodic and non-deterministically random) systems within the framework of the
modern theory of the insulating state \cite{Kohn, RS, SWM, RestaTopical, Resta}. We discern metallic and insulating phases via the many-particle localization tensor which signals the 
so-called Kohn's localization in the many-particle ground-state~\cite{Kohn, Resta, SWM, Bendazolli, Varma}. We highlight the important difference between the conduction properties predicted via 
Kohn's many-particle localization tensor to those evinced from 
the analysis of other measures of localization familiar from the theory of Anderson insulators, such as the participation ratio or the Lyapunov exponent, which capture the localization properties of the 
single-particle eigenstates.

The first system investigated is the off-diagonal Fibonacci chain, where the 
quasi-periodicity is present in the hopping terms dictated by a substitution rule, in turn determined by the Fibonacci sequence. 

The second model is the Riemann chain, in which the lattice 
spacings are defined from the (renormalised) distances between the nontrivial zeros of the Riemann zeta function. We introduce this model inspired by Dyson's proposal on a possible strategy to 
prove or disprove the Riemann hypothesis. According to Dyson \cite{Dyson}, if one were able to identify the Riemann zeta-function zeros with the nodes of a one-dimensional quasicrystal, 
then the Riemann conjecture would be proved (see Sec. \ref{section3} for details). 
It, therefore, behoves us to investigate the conduction properties of the Riemann chain, and to compare them to those of a paradigmatic quasicrystal (Fibonacci chain) and to those of random Anderson models 
with non-deterministic disorder.

%%%%%%%%%%%%%%%%%%%%%%%%%%%%%%%%%
%%%%%%%%%%%%%%%%%%%%%%%%%%%%%%%%%%%%
%%%%%%%%%%%%%%%%%%%%%%%%%%%%%%%%%%%%%
A primary finding of our work is that the bulk conductivity (or insulation) in these two models strongly depends on the lattice filling. 
In the Fibonacci chain, Kohn's many-particle localization tensor is divergent in the thermodynamic limit at most values of the filling, 
including the half-filled case. This indicates that the Fibonacci lattice is, in general, a metallic system. However, at certain special filling factors, Kohn's many-particle localization length 
remains finite in the thermodynamic limit, signalling an insulating phase. The origin of these insulating points is discussed.
Instead, in the Riemann lattice, the system appears to be insulating at all fillings, excluding only the half-filled case. This behaviour is unlike that of the Fibonacci quasicrystal, 
and is instead similar to that of the off-diagonal Anderson model with random hoppings (in contrast to Dyson's proposal taken verbatim). Still, some differences emerge also with respect to the case of 
random disorder, specifically in the anomalies displayed by both models at half filling. Therefore, we refer to the Riemann lattice as a quasi-random model.

The remainder of the article is organized as follows:
 in Section~\ref{sec1} we review the basic concepts of the modern theory of the insulating state, describing in particular the connection between Kohn's many-particle localization tensor and 
 the dc conductivity, and  we also discuss other measure of localization commonly employed in the theory of Anderson insulator.
 In Sections~\ref{section2} and ~\ref{section3} we discuss the results for the Fibonacci and the Riemann lattices, respectively.
 A summary of the results, and a critical discussion on the power of the modern theory of the insulating state, are reported in Section~\ref{Conclusions}.
 Appendix~\ref{appendix1} reports a detailed analysis on the system-size scaling of Kohn's many-particle localization tensor, while in appendix~\ref{appendix2} the diffraction pattern of the Fibonacci chain 
 is described, and the cut-and-project method to create the Fibonacci chain is illustrated. Appendix \ref{appendix3} describes the multifractal analysis employed for the Fibonacci quasicrystal when comparing 
 with the expected scaling of the participation ratio.
%%%%%%%%%%%%%%%%%%%%%%%%%%%%%%%%%%%%%%%%%%%%%%%%%%%%%%%%%%%%%%%%%%%%%%%%%%%%%%%%%%%%%%%%%%%%%%%%%%%%%%%%%%%%%%%%%%%%%%%
%%%%%%%%%%%%%%%%%%%%%%%%%%%%%%%%%%%%%%%%%%%%%%%%%%%%%%%%%%%%%%%%%%%%%%%%%%%%%%%%%%%%%%%%%%%%%%%%%%%%%%%%%%%%%%%%%%%%%%%
%%%%%%%%%%%%%%%%%%%%%%%%%%%%%%%%%%%%%%%%%%%%%%%%%%%%%%%%%%%%%%%%%%%%%%%%%%%%%%%%%%%%%%%%%%%%%%%%%%%%%%%%%%%%%%%%%%%%%%%
%%%%%%%%%%%%%%%%%%%%%%%%%%%%%%%%%%%%%%%%%%%%%%%%%%%%%%%%%%%%%%%%%%%%%%%%%%%%%%%%%%%%%%%%%%%%%%%%%%%%%%%%%%%%%%%%%%%%%%%
%
\section{Model and method}
\label{sec1}
We consider a system of $N/2$ spin-up and $N/2$ spin-down noninteracting electrons on an open chain \cite{BoundNote} of $L$ sites. 
The system is described by the following tight-binding Hamiltonian:
\begin{eqnarray}
\label{eq: BasicHamiltonian}
  H = \sum_{{r,\sigma}}t_r (b_{{r,\sigma}}^{\dagger}b^{\phantom{\dagger}}_{{r}+{1},\sigma} + 
  \textrm{h.c})
 \raggedleft
 \label{hamiltonian}
 \end{eqnarray}
 where ${r}=1,\dots,L-1$ is the discrete index which labels the lattice sites,  
 $b_{r,\sigma}$ ($b_{r,\sigma}^{\dagger}$) is the fermionic annihilation (creation) operator for a 
 spin $\sigma=\uparrow,\downarrow$ particle at site $r$. 

The Hamiltonian \eqref{eq: BasicHamiltonian} possesses a sub-lattice chiral symmetry. This means that $H$ may be decomposed into off-block diagonal form:
\begin{equation}
 \label{eq: OffBlock}
 H = \left( \begin{array}{cc}
0 & h  \\
h^{\dagger} & 0 \end{array} \right),
\end{equation}
where $h$ is the hopping matrix connecting odd and even sites, and the chiral symmetry, given by 
\begin{equation}
 \label{eq: Csymmetry}
 \tilde{\sigma}_z H \tilde{\sigma}_z = -H ,
\end{equation}
holds. Here $\tilde{\sigma}_z = \mathbbm{1}_{L/2}  \otimes\sigma_z $, where $\sigma_z$ is the third Pauli matrix, and $ \mathbbm{1}_{L/2}$ is the $L/2 \times L/2$ identity matrix. 

In the theory of the  Anderson transition developed for randomly disordered models, one discerns metallic behaviour from insulation by inspecting the spatial extent, respectively ergodic or localized,  
of the single-particle eigenstates at the Fermi energy. This spatial extent can be analysed by computing the participation ratio (PR), which is defined by:
\begin{equation}
 \label{eq: PR}
 \textrm{PR}_{j} = 1/\sum_{r} |\phi_{j}({r})|^4,
\end{equation}
where $|\phi_{j}(r)|$ is the absolute amplitude of the eigenstate labelled by $j$ at site $r$. For a localized state, the PR is independent of $L$, 
whereas for ergodic delocalized states it scales with the system size as $\textrm{PR}_{j} = \mathcal{O}(L^d)$, where $d$ is the dimensionality, meaning that the PR diverges in the thermodynamic limit. 
For critical states $\textrm{PR}_{j} = \mathcal{O}(L^{b})$, with $0 < b < d$.

Another measure of localization of the single-particle eigenstates which, however, is known not bear a one-to-one correspondence with the PR  \cite{Yudson, Johri}, is the inverse Lyapunov exponent $\xi_1$.
For a single-particle state $\phi (r)$ it is defined as:
\begin{equation}
 \label{eq: LyapunovLL}
 1/\xi_1 = -\lim_{r \rightarrow \infty} \frac{\log{|\phi(r)|}}{r}.
\end{equation}
This measure captures only the properties of the tail of the wavefunction and can therefore entirely miss bulk properties. In particular, if the wavefunction is sub- or superlocalized (i.e. stretched 
exponential decay) then the definition in Eq. \eqref{eq: LyapunovLL} may prove to be inadequate. Indeed, this is the case for the off-diagonal Anderson model as we explain in Sec. \ref{section3}. 

A primary theoretical tool we employ in this article in order to investigate the bulk conduction and insulation properties of the aforementioned models is
the many-particle localization tensor $\lambda$ defined within the modern theory of the insulating state \cite{RS, Resta, SWM, RestaTopical}. 
This localization tensor $\lambda$ is a property of the many-particle ground-state wave-function, and it is related to the fluctuation of the polarization. 
In previous studies $\lambda$ has proven to be suitable to identify various insulating 
phases, including band, Anderson \cite{Varma,RestaNEW}, Mott \cite{RS}, and also quantum Hall insulators \cite{RestaQHP}. 
Furthermore the effect of weak interactions on the Anderson transition has also been addressed \cite{Varma}.
The absence of conduction as signalled by an $L$-independent $\lambda$ is referred to as \textit{Kohn's localization}; it reflects the localization in the ($dN$-dimensional) configuration space 
\cite{Resta, SWM, Bendazolli, Varma}, as originally discussed by  Kohn in his seminal article on the theory of the insulating state~\cite{Kohn}.
An important difference between $\lambda$ and the previously discussed measures of localization such as PR and Lyapunov exponent is that \emph{ $\lambda$ captures spectral properties as well.} 
This makes it suitable for systems with complex spectra, such that those with a heirarchy of mini-gaps.
In this article, we analyse this metric for localization of the many-particle ground state of certain aperiodic one-dimensional models, whose properties are intermediate between clean periodic systems and 
random models with non-deterministic disorder.

                     \begin{figure*}[ttp!]
\centering
\includegraphics[scale=0.95]{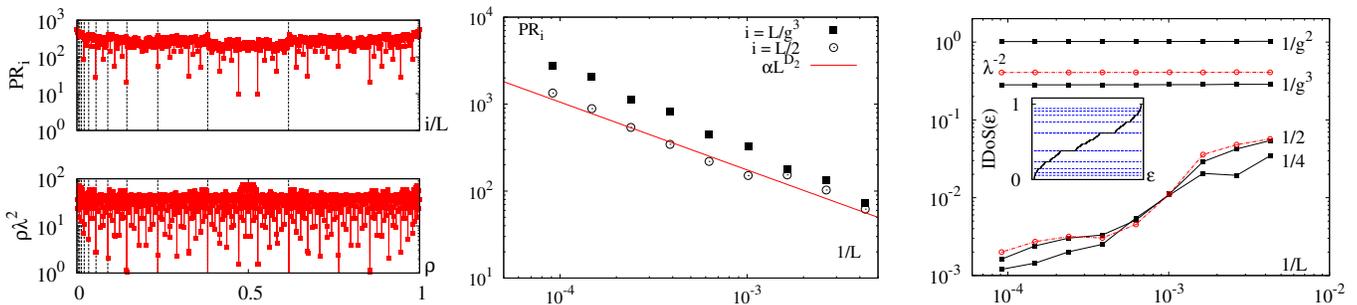}
     \caption{(Colour online) Conduction properties in the Fibonacci quasicrystal. 
      Left top panel: participation ratio (PR) of the single-particle wave-functions as a function of the scaled eigenstate index $i/L$, for system size $L=988$, on a log-normal scale.
     All single particle states are critical in the Fibonacci lattice
     \cite{Sutherland, Banavar, Fujiwara, Hiramoto} in the thermodynamic limit.
     Left bottom panel: rescaled (squared) many-particle localization length $\rho\lambda^2$ as a function of filling $\rho$ on a log-normal plot. The vertical dashed lines correspond to the location of the 
     Fibonacci numbers; at these fillings, $\lambda^2$ shows a sudden dip, 
     indicative of an insulating phase. 
     Middle panel: Scaling of PR with inverse system size for eigenstate numbered $i$ located at the beginning of a mini-gap and at the band centre.
     Solid line shows the expected scaling with a generalised dimension $D_2 \approx 0.78$ for the multifractal state at band centre for $q=2$ (corresponding to the PR), obtained from a 
     multifractal analysis \cite{Chhabra, Schreiber, Grussbach}, described in Appendix \ref{appendix3}.
     Right panel: Scaling of $\lambda^{-2}$ as a function of inverse system size for fillings 
     $\rho = 1/4, 1/2, 1/g^2, 1/g^3$. 
     The suppression of $\lambda^2$ in the thermodynamic limit for the former two densities clearly indicates a metallic phase, 
     whereas the saturation at electronic fillings of $1/g^n$ signals a many-particle insulator. Dot-dashed (red circled) lines correspond to $t_2/t_1 = 0.5$, whose results are qualitatively unchanged.
     Inset shows the integrated density of states across the energy spectrum (horizontal axis) of an 
     $L=2585$ chain; it is worth noting that at the values $1/g^n$, indicated by dashed horizontal lines (which correspond to the black vertical lines in the left panel), 
     there is also a gap in the spectrum, signalled by the presence of plateaus. Instead, the PR values cannot signal these insulating densities as seen in the middle panel.
     }  
     \label{fig: F5}
   \end{figure*}
   \vspace{-0.5cm}
\subsection*{Insulation vs. localization}
%%%%%%%%%%%%%%%%%%%%%
%%%%%%%%%%%%%%%%%%%%%%%
In the case of noninteracting electrons, which we consider here, the ground state wavefunction $\Psi$ is constructed using Slater determinants built from the single-particle spatial orbitals denoted by 
$\phi_j(\textbf{r})$, for $j=1,2, \ldots L$.
For a Slater determinant wave-function, the localization tensor may be evaluated as \cite{RS, Resta, Varma}
\begin{equation}
 \label{eq: locOpen}
 \lambda^2_{\alpha \beta} = \frac{1}{N}\int d\textbf{r}d\textbf{r}'(\bf{r} - \textbf{r}')_{\alpha}(\textbf{r} - \textbf{r}')_{\beta}|P(\textbf{r},\textbf{r}')|^2,
\end{equation}
where $\alpha, \beta$ correspond to spatial coordinates; $\rho_D(\textbf{r}, \textbf{r}') = 2P(\textbf{r},\textbf{r}')$ 
is the one-particle density matrix for a Slater determinant, 
which in turn is given by \cite{Resta} $\rho_D(\textbf{r}, \textbf{r}') = 2\sum_{j=1}^{N/2}\phi_j(\textbf{r})\phi^{*}_j(\textbf{r}')$. 
The single-particle spatial wave-functions $\phi_j(\textbf{r})$, needed to form the  
one-particle density matrix $\rho_D$ of the many-particle system, are determined from full diagonalisation of the Hamiltonian 
matrix for a single particle using the Armadillo library \cite{Armadillo}{}. 

The length-scale $\lambda$ (we suppress $\alpha, \beta$ subscripts from here on, since we deal with one-dimensional models) is a many-particle localization length determining Kohn's localization of the 
ground-state of the many-particle system. 
The scaling of $\lambda$ with system size allows one to distinguish a conductor from an insulator \cite{RS, SWM, Varma}; for sufficiently large size, $\lambda$ diverges with $L$ if the zero-frequency conductivity $\sigma(\omega=0)$ is finite (i.e., for metals), 
while it saturates to a finite value if  $\sigma(\omega=0) = 0$ (i.e., for insulators).
%%%%
This scaling behaviour can be evinced from the following sum-rule which relates Kohn's many-particle localization length to the frequency-dependent conductivity $\sigma(\omega)$  \cite{RS, SWM, RestaOBC}: 
\begin{equation}
 \label{eq: FlucDiss}
 \lambda^2 = \frac{\hbar}{\pi e^2\rho} \int_{0}^{\infty}\frac{\textrm{d}\omega}{\omega}\sigma(\omega).
\end{equation}
This fluctuation dissipation relation is valid for noninteracting electrons with any boundary conditions \cite{RestaOBC} 
and for generic interacting systems with periodic boundaries \cite{SWM, Resta},
the former scenario being pertinent to our study. 
By considering the generalized Einstein relation at low temperature \cite{Imry, Kubo}, that is 
\begin{equation}
 \label{eq: Einstein}
 %\sigma (\omega \rightarrow 0) = e^2 D(\epsilon) \frac{d }{d \epsilon}(\textrm{DoS}(\epsilon)),
 \sigma (\omega \rightarrow 0) = e^2 \textrm{DoS}(\epsilon) D(\epsilon),
\end{equation}
where $\textrm{DoS}(\epsilon)$ is the density of states (see Eq. \eqref{eq: DOS}) close to the Fermi energy and $D(\epsilon)$ is the diffusion constant at energy $\epsilon$,
one readily understands that $\lambda$ has to be sensitive to both spectral properties such as singularities or gaps in the DoS,
as well as the transport properties of the single-particle eigenstates, which are reflected through the diffusion constant $D(\epsilon)$.
These expectations will be borne out by our numerical results.

In most physically relevant scenarios, the functional form of the system-size scaling of $\lambda$ can be predicted.
Let us consider the cases of insulators and metals separately. For insulation due to single-particle eigenstate localization in $d$-dimensions, it may be shown (see Appendix \ref{appendix1}) that for 
power-law or stretched-exponential localization, with $\mu > d$, the low-frequency conductivity scales as:
\begin{equation}
\label{eq: Condscaling}
\sigma(\omega \rightarrow 0) \sim 
  \begin{cases} 
   \omega ^ {2 - (d+1)/\mu} & \text{if } |\phi |^2 \sim R^{-\mu} \\
   \omega^2 \log^{(d+1)/\alpha}{(\omega)}       & \text{if } |\phi |^2 \sim \exp{(-b R^{\alpha})}.
  \end{cases}
\end{equation}
This will in turn imply a saturation in $\lambda^2$ as $L \rightarrow \infty$, see Appendix \ref{appendix1}. Saturation of $\lambda^2$ when there is a low-frequency optical gap may 
be argued by upper-bounding the value of the integral in Eq. \ref{eq: FlucDiss} \cite{SWM, Resta}.
In the metallic case, the d.c. conduction in the thermodynamic limit is finite: $\sigma(\omega \rightarrow 0) \neq 0$; then, it can be shown that $\lambda$ will pick up factors of $\log{L}$ in its scaling 
(for diffusive), along with factors of $L$ (for ballistic); see Appendix \ref{appendix1}.

We point out that due to the symmetry of \eqref{eq: BasicHamiltonian} $\rho\lambda^2 \equiv \mathcal{K}(L,\rho)$
will be a function $\mathcal{K}$ that is symmetric about half-filling i.e. $\mathcal{K}(L,\rho) = \mathcal{K}(L,1-\rho)$ 
where $\rho = N/2L$.  In a clean tight-binding model or in the Aubry-Andr\'{e} model in the conducting phase $\mathcal{K}(L,\rho) = \mathcal{K}(L)$ (i.e. independent of the density) 
with $\mathcal{K}(L)$ increasing with system size $L$. 
In comparison, we shall observe very different behaviour for the function $\mathcal{K}$ in the three systems under consideration.

%
%%%%%%%%%%%%%%%%%%%%%%%%%%%%%%%%%%%%%%%%%%%%%%%%%%%%%%%%%%%%%%%%%%%%%%%%%%%%%%%%%%%%%%%%%%%%%%%%%%%%%%%%%%%%%%%%%%%%%%%
%%%%%%%%%%%%%%%%%%%%%%%%%%%%%%%%%%%%%%%%%%%%%%%%%%%%%%%%%%%%%%%%%%%%%%%%%%%%%%%%%%%%%%%%%%%%%%%%%%%%%%%%%%%%%%%%%%%%%%%

\section{Fibonacci lattice} 
  \label{section2}
  The spectral properties of Fibonacci quasicrystals have been the subject of various theoretical and experimental 
studies since the 1980's \cite{Kadanoff, Siggia, MerlinExpt}. In Ref. \citen{MerlinExpt} the authors experimentally created a 
Fibonacci quasicrystal from alternating layers of GaAs and AlAs; X-ray and Raman scattering analysis revealed the presence of 
singularities in the spectrum. 
In subsequent theoretical works \cite{Sutherland, Banavar, Fujiwara, Hiramoto} the spectral 
properties of the Fibonacci lattice have been analysed, establishing that all eigenstates (both in the diagonal case where the on-site energies are modulated, and in the off-diagonal 
case where the modulation is in the hopping energies) are critical and display multifractal properties; the spectrum was found to be purely singular continuous. 
The conduction properties have been addressed, but mostly within the Landauer formalism. Results consistent with a power-law growth of the Landauer resistance with systems size have been reported in 
Refs.~\citen{SutherlandKohmoto, Sarma}, with large fluctuation depending on the energy~\cite{Sarma}. The presence of states displaying the transport properties typical of extended states has also been reported~\cite{Macia}.
On the experimental side, a proposal was recently put forward on how to realise the Fibonacci quasicrystal in experiments performed with ultracold atomic gases by employing a narrow-width confining Gaussian beam on a 
square two-dimensional optical lattice \cite{Singh}.

The Fibonacci quasicrystal is constructed using a simple replacement rule of two symbols $L, S$:
\begin{eqnarray}
 \label{eq: FiboRule}
 \{S\} \rightarrow \{L\}, \nonumber \\
 \{L\} \rightarrow \{LS\}.
\end{eqnarray}
The finite sequences generated will then be $\{S, L, LS, LSL, LSLLS, LSLLSLSL \ldots\}$. The transformation matrix that generates this sequence is given by 
\begin{equation}
 \label{eq: Trans}
 M =  \left(
 \begin{array}{cc} 
 1 & 1 \\ 1 & 0 
 \end{array} \right),
\end{equation}
whose eigenvalues are given by $g = (\sqrt{5} + 1)/2$ and  $1/g$. $g$ is a Pisot-Vijayaraghavan number and the sequence generated by $M$ is therefore a valid quasicrystal \cite{Senechal}.
We construct a Fibonacci chain consisting of points with the two bond lengths $\delta_{r}=\{1,g\}$ (these two values emerging from the cut-and-project method, see ~\ref{appendix2} for more details) arranged in the Fibonacci sequence \cite{Steinhardt, Senechal}. 
It will correspond to two hopping values $t_{1}$ and $t_{2}$ arranged in the same sequence. 
For concreteness we consider the following correspondence $t_{r}=1/\delta_{r}$ between the structure of the quasi-crystal (encoded in $\delta_{r}$) and the tight-binding Hamiltonian. 
In this case $t_{1}=1$ and $t_{2}=1/g$. We will demonstrate that the main qualitative results are insensitive to the particular relation between $t$ and $\delta$ and 
result only from the Fibonacci sequence of $t_{1}$ and $t_{2}$. 
The length of each such sequence (number of bonds in the lattice) is a Fibonacci number $F_i$, 
which will in turn determine the length of the lattices $L \equiv F_i + 1$ that may be studied.

We first analyse the PR of the single-particle eigenstates. 
The results are displayed in the left top panel of 
Fig. \ref{fig: F5}. We display data corresponding to the entire spectrum for the system size $L=988$. The horizontal axis indicates the eigenstate index divided by the corresponding system size.
Several sharp dips in the PR values are evident. These correspond to more localized states; the ``ceiling'' from which these dips hang move upward as the system size increases (not shown), corresponding to a tendency to delocalization. 
\textit{Note that these states are neither truly localized nor ergodic states; in the thermodynamic limit all states are expected to be critical} \cite{Sutherland, Banavar, Fujiwara, Hiramoto}. 
This critical nature is indeed confirmed by the system-size scaling analysis of the PR  values displayed in the middle panel of Fig. \ref{fig: F5}. We consider in particular two eigenstates: one just below a mini-gap and one at the band-centre. 
In both cases a power-law scaling
$\textrm{PR} \propto L^{D_2(\epsilon)}$ consistent with multifractality, with $D_2(\epsilon) < d$ is seen [we recall that one would have $D_2=0$ for localized states, and $D_2=d=1$ for ergodic states], 
where $\epsilon$ is the energy of the eigenstate under consideration and $d$ is the dimensionality of the lattice; 
the full line shows the expected scaling at $\epsilon=0$, with the generalised dimension $D_2(\epsilon = 0)$, which we obtained from a multifractal analysis \cite{Chhabra, Schreiber, Grussbach} 
(see Appendix \ref{appendix3} for details).

\begin{figure*}[ttp!]
            \hspace*{-0.8cm}
\centering
\includegraphics[scale=0.5]{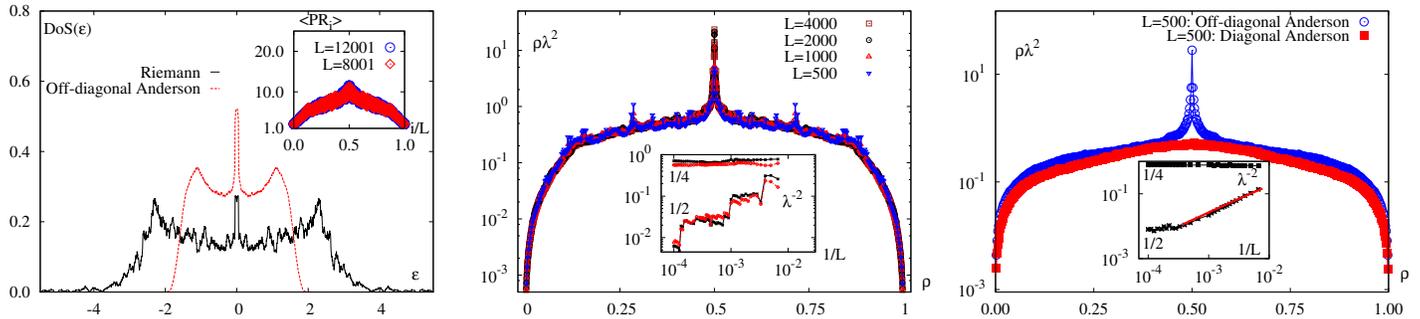}
     \caption{(Colour online) Conduction properties in the Riemann lattice and Anderson models. 
     Left panel: Density of states of the Riemann lattice and disorder-averaged off-diagonal Anderson model. There is a sharp rise in the density of states towards the band-centre 
     $\epsilon=0$, more prominently seen in the off-diagonal Anderson model. System size $L=4000$ is chosen, with open boundaries.
     Inset shows participation ratio ($\textrm{PR}$) of the single-particle eigenstates for various system sizes $L$ as a function of the scaled eigenstate index $i/L$ in the Riemann lattice; 
     $<>$ denotes averaging of inverse participation ratio over 50 neighbouring eigenstates. 
     Middle panel: $\rho\lambda^2$ for various system sizes in the Riemann lattice as a function of electronic 
     filling $\rho$, on a log-normal scale. The sharp increase at half-filling, associated to the chiral symmetry $\sigma_z H \sigma_z = -H$, indicates the occurrence of a metallic phase.
     Inset shows scaling of $\lambda^{-2}$ as a function of inverse system size, with exponential (red circles) and algebraic (black crosses) relation between the spacings $\delta_r$ and hoppings $t_r$, 
     for two electron fillings $\rho = 1/4, 1/2$. In the first case,  $\lambda$ saturates in the thermodynamic limit, signalling an insulator, in the second, $\lambda$ diverges in the thermodynamic limit, 
     confirming the occurrence of the metallic phase at half filling.
     Right panel: Disorder-averaged $\lambda^2$ for the open boundary off-diagonal and the diagonal Anderson models.
     Note the similarity of the former with the Riemann lattice (due to chiral symmetry) and the absence of a sharp peak at half-filling in the latter (where chiral symmetry is absent).
     Inset shows scaling of $\lambda^2$ for the off-diagonal Anderson model for $\rho = 1/4, 1/2$; the full red line is a fit to $(L \log{L})^{-1}$ (see text) signalling quasi-ballistic transport at small 
     system sizes. 
     }  
     \label{fig: F3}
   \end{figure*}

In order to ascertain the conducting properties of the noninteracting many-particle system, whose wave-function is the Slater determinant composed of these critical wavefunctions, 
we analyse Kohn's many-particle localization tensor $\lambda$.
The dependence of $\rho \lambda^2$ as a function of the filling $\rho$ is displayed in the bottom left panel of 
Fig. \ref{fig: F5}, for  the lattice size $L= 988$.
In general, the variations of $\lambda^2$ as a function of $\rho$ display very sharp features; this behaviour is to be contrasted with the smoother dependence we will observe in case of the Riemann lattice
 and Anderson models (Fig. \ref{fig: F3} of next section); we attribute this to the critical nature of the single-particle wavefunctions and to the presence of small gaps in the Fibonacci quasicrystal. 
At various specific densities, we observe sharp dips in the localization length; this indicates a decrease in conducting properties, possibly the onset of insulating behaviour (see scaling analysis below). 
Several, but not all, of these specific densities are given by the relation $F_i/L$ where $F_i$ is the $i^{\textrm{th}}$ Fibonacci number. 
Notice that these values can be approximated by $1/g, 1/g^2, 1/g^3 \ldots$). In Fig. \ref{fig: F5}, these densities are indicated by vertical lines.
This pattern in the localization properties is consistently reproduced for different system sizes, meaning that the localization dips obtained for different lattice lengths occurs at the same fillings.
We have also checked that a very similar structure is obtained for different values of the ratio between the two hopping energies (we considered values varying over a few orders of magnitude).

In order to verify the above statement about possible insulating behaviour at the location of the sharp dips --- in particular those corresponding to the vertical lines --- 
we analyse the scaling of the localization length $\lambda$  with the systems size, keeping the electronic density fixed. The cases of the densities $\rho = 1/g^2$ and $\rho= 1/g^3$ are shown in the 
right panel of Fig. \ref{fig: F5} (notice that particle filling of $1/g$ is equivalent to $1/g^2$ because of the symmetry about half-filling; this is, in fact, evident from Fig. \ref{fig: F5}). 
We observe that $\lambda$ quickly saturates to finite values as the system size increases, clearly indicating an insulating state.
The origin of these insulating ``dips'' can be understood from the analysis of the single-particle spectrum. To elucidate this point, in the inset of the right panel of Fig. \ref{fig: F5} we show 
the integrated density of states $I(\epsilon)$ as a function of the energy $\epsilon$ (for an $L=2585$ chain). The plateaus in $I(\epsilon)$ correspond to the gaps in the single-particle energy spectrum. 
Such a structure of the spectrum, often referred to as the devil's staircase, is 
typical of cumulative distributions of a Cantor function and was previously observed in the Fibonacci quasicrystal \cite{Luck, Luck2}. 
We see that the plateaus in the cumulative density of states $I({\epsilon})$ correspond to the (normalized) integrated density of states at $1/g^n$ (as previously reported \cite{Luck2}); 
this means that the insulating dips described above occur exactly at the densities where the Fermi energy is at the verge of a band gap.

For generic densities (away from the sharp dips, i.e. the mini-gaps), the conduction properties of the electron gas in the Fibonacci quasicrystal are more enigmatic, since the single-particle 
eigenstates are critical.
To clarify this, we analyse the finite-size scaling behaviour of $\lambda$. 
We display in particular two cases, namely half filling $\rho = 1/2$ and quarter filling $\rho/1/4$ (see right panel of Fig.~\ref{fig: F5}).
$\lambda$ clearly diverges in the thermodynamic limit, indicating that the Fibonacci quasicrystal is, for these two densities, a metallic system. We verified a similar divergence for other generic 
filling away from the mini-gaps.

%%%%%%%%%%%%%%%%%%%%%%%%%%%%%%%%%%%%%%%%%%%%%%%%%%%%%%%%%%%%%%%%%%%%%%%%%%%%%%%%%%%%%%%%%%%%%%%%%%%%%%%%%%%%%%%%%%%%%%%
%%%%%%%%%%%%%%%%%%%%%%%%%%%%%%%%%%%%%%%%%%%%%%%%%%%%%%%%%%%%%%%%%%%%%%%%%%%%%%%%%%%%%%%%%%%%%%%%%%%%%%%%%%%%%%%%%%%%%%%
   %\vspace{-0.5cm}
  \section{Riemann lattice}
    \label{section3}
  The Riemann zeta function is one of the most studied functions in number theory. It was defined by Riemann in his seminal article \cite{Riemann} 
  ``Ueber die Anzahl der Primzahlen unter einer gegebenen Gr\"{o}sse'' as 
  \begin{equation}
   \label{eq: RZF}
   \zeta(s) = \sum_{n=1}^{\infty}\frac{1}{n^s},
  \end{equation}
  for a complex variable $s$. The nontrivial zeros of $\zeta(s)$ are conjectured to all lie on the critical line $1/2 + i\gamma_n$, with real valued $\gamma_n$, with $n \in \mathbb{Z}$. This 
  statement constitutes the Riemann hypothesis \cite{Curiosity}. The distribution of the values $\left\{\gamma_n\right\}$ has been intensively studied; one such  
  possible structure in the imaginary part of the zeros is by Dyson who conjectured a possible connection between quasicrystals and the 
  nontrivial zeros of $\zeta(s)$ \cite{Dyson}: per Dyson, if the set $\left\{\gamma_n\right\}$ forms a one-dimensional quasicrystal, then the Riemann hypothesis is proved;
  his conjecture is based on the definition of quasicrystals Eq. \eqref{eq: RiemannFourier} wherein a discrete set of real-space positions gives another discrete set of well-defined diffraction points 
  upon Fourier transforming. 
  More formally, Dyson claims \cite{Dyson} that, following our notation of Eq. \eqref{eq: RiemannFourier}, 
  \begin{equation}
   \label{eq: PrimeDyson}
   k_{m} = \log{p^m},
  \end{equation}
   with $X^{*}$ containing all primes $p$ and integers $m$, and that therefore Eq. \eqref{eq: RiemannFourier} is satisfied provided \textit{all} $\gamma_n$ are real, which is equivalent to assuming 
   the validity of the Riemann hypothesis.
   
  %%%
  While certain objections to Dyson's proposal may be raised \cite{VarmaFuture},
  we pursue its physical implications by defining a lattice with points located at the values $\gamma_n > 0$. We refer to this model as the \textit{Riemann lattice}. 
  We note that the zeros of $\zeta(s)$ become denser as one traverses higher up the critical line: the average spacing between consecutive zeros at a given 
  height $z$, for $z \rightarrow \infty$, on this critical line is given by \cite{Odlyzko}
    \begin{equation}
   \label{eq: LargeZeros}
   N(z) = 2\pi/\log(z/2\pi).
  \end{equation} 
  Therefore, in order to construct a model with an average lattice spacing of unity, we utilise the gap $\delta_r$ between the renormalised zeros 
  defined by \cite{Odlyzko}
  \begin{equation}
   \label{eq: RenormRZ}
   \delta_r = (\gamma_{r+1} - \gamma_r)\log{(\gamma_r/2\pi)}/2\pi.
  \end{equation}
  The hoppings $t_r$ in Eq. \eqref{eq: BasicHamiltonian} are taken to be (i) $t_r = 1/\delta_r$, and (ii) $t_r = \textrm{exp}(-\delta_r)$; the physics of the Riemann lattice is 
  qualitatively unchanged between the two choices.
  With the above transformation the average spacing between the renormalised zeros is unity in the sense that \cite{Odlyzko}
  \begin{equation}
   \label{eq: RZspacing}
   \sum_{r=j+1}^{j+k}\delta_r = k + O(\log{(jk)}).
  \end{equation}
 Below, we analyse the bulk conduction properties of the Riemann lattice, and we compare them to those of a quasiperiodic model such as the Fibonacci quasicrystal 
 [discussed in Section~\ref{section2}], and to those of a random model such as the Anderson models with non-deterministic disorder.
 
The Hamiltonian defining the disordered Anderson models \cite{Anderson} is the following:
  \begin{equation}
   \label{eq: Anderson}
   H_{\textrm{And.}} = H + \sum_{\sigma, r=1}^L \epsilon_{r,\sigma} b^{\dagger}_{r,\sigma}b^{\phantom{\dagger}}_{r+1,\sigma};
  \end{equation}
$H$ is given by Eq. \eqref{eq: BasicHamiltonian}, and $t_r$ and $\epsilon_{r,\sigma}$ are uniformly distributed non-deterministic random variables. 
Specifically, we consider the off-diagonal Anderson model with hoppings $t_r \in [0.1,1.1]$ and no onsite disorder$\epsilon_{r,\sigma}=0$, and also the diagonal Anderson model with 
$\epsilon_{r,\sigma} \in [-2,2]$  and constant hoppings $t_r =1$. 
In the off-diagonal Anderson model there is a singularity in the \textit{average} density of states at the band-centre $\textrm{DoS}(\epsilon) \propto 1/(\epsilon \ln^3{\epsilon})$ 
\cite{Dyson2, Inui, McKenzie}. This singularity is associated with the chiral symmetry discussed in Section~\ref{sec1}~\cite{Dyson2}. According to the Herbert-Jones formula \cite{HerbertJones}, the single-particle Lyapunov 
localization length at energy $\epsilon$ is related to the DoS by 
$\xi_1^{-1}(\epsilon) = \int^{\infty}_{-\infty}\textrm{DoS}(E)\ln{|\epsilon - E|}\textrm{d}E$.
This implies a logarithmic divergence of the single-particle 
Lyapunov localization length $\xi_1 \propto \ln{\epsilon}$ in the zero-energy limit $\epsilon\rightarrow0$, as has been confirmed numerically \cite{Biswas}. 
However, this divergence may be an artefact of the assummed exponential functional form of the decay of the single-particle wave-function $\phi(r) \sim \exp{(-r/\xi_1)}$. 
In fact, Refs. \citen{Fleishman, Inui} argue for sublocalization $\phi(r) \sim \exp{(-\sqrt{r/\xi_1})}$ rather than anomalous delocalization, 
and in Ref.~ \citen{Inui} it has been found that that the \textit{typical} localization length (in contrast to the average localization length) is finite. Furthermore, it has been shown that due to large 
fluctuations in $\xi_1$ \cite{Fleishman} and to the algebraic decay of the \textit{average} transmission coefficient 
with the system size \cite{Economou} $T \sim L^{-\gamma}, \gamma \approx 0.5$, transport in the $\epsilon = 0$ state is inhibited. 
It is important to mention that the PR values do not display any anomalous peak (which would imply a large or divergent single-particle localization length) at the band centre~\cite{Kozlov}, 
confirming that there is no exact correspondence between Lyapunov localization length and PR values \cite{Yudson, Johri}.
In the diagonal Anderson model, the band-centre anomaly is present but both single-particle localization lengths are always finite across the spectrum \cite{Yudson}.
%
%\textit{Localization metrics}:

   We begin the analysis of the Riemann lattice by studying its single-particle properties. 
   The density of states at energy $\epsilon$ defined as 
   \begin{equation}
    \label{eq: DOS}
    \textrm{DoS}(\epsilon) = \sum_n \delta_{\textrm{D}}(\epsilon - \epsilon_n),
   \end{equation}
where $\epsilon_n$ is the eigenenergy of the level labelled by $n$, is shown in the left panel of Fig. \ref{fig: F3}. 
A sharp peak at the band centre is evident. Such a peak is present also in the case of the off-diagonal Anderson model (also shown in Fig. \ref{fig: F3}), as discussed above.

   The second contributing factor to $\sigma(\omega \rightarrow 0)$ per Eq. \eqref{eq: Einstein} is the diffusion constant $D(\epsilon)$ which we infer through the PR, 
   whose averaged value across the spectrum is shown in the inset of Fig. \ref{fig: F3}. While there is a generic delocalizing effect towards the band-centre (as is the case even for fully localized 
   models such as the 1D diagonal Anderson model \cite{Anderson, Gang4, Varma}), PR of single-particle eigenstates at or close to the band-centre show no clear scaling as $L \rightarrow \infty$ (not shown) 
   in contrast to what we observed for the Fibonacci quasicrystal.
   A more conclusive statement about dc conductivity may be made from the results of $\lambda^2$ displayed in the middle panel of Fig. \ref{fig: F3}, 
   as we describe next.
   
   The most important observations that may be drawn from this figure are the following: (i) the numerical coincidence of $\lambda^2$ for various $L$ at almost all values of filling $\rho$, possibly indicating an 
   insulator, and (ii) an anomalous increase in $\lambda^2$ at half-filling with its values seemingly increasing with $L$, possibly indicating a conductor.
   These two points (general insulating behaviour, and metallicity only at half-filling) are confirmed by fixing $\rho = 1/4, 1/2$ and approaching the thermodynamic limit as displayed in the inset: 
   $\lambda^2$ saturates in the former case and diverges in the latter. We attribute the enhanced conductivity in the half-filled case to the peak in the DoS.
   
   Let us compare and contrast this situation to other systems. The behaviour is markedly different from the cases of the Fibonacci quasicrystal (which is conducting at almost all densities excepting 
   those finely-tuned to the gaps) and the diagonal Anderson model (which is insulating at all densities \cite{Anderson, Gang4, Varma} and has no band-centre anomaly in $\lambda^2$, as seen in right panel of 
   Fig. \ref{fig: F3}). However it is analogous to the off-diagonal Anderson model, both of which are insulating at all densities except at $\rho=1/2$ where an anomalous increase in $\lambda^2$ is observed 
   (middle and right panel of Fig. \ref{fig: F3}). Indeed comparing the insets of these panels we see that while the systems at $\rho=1/4$ are clearly insulators, the half-filled cases are 
   more subtle: in the off-diagonal Anderson an initial scaling $\lambda^2 \propto L\log{L}$ (corresponding to quasi-ballistic transport, see Appendix \ref{appendix1}) 
   crosses over to either a saturation in $\lambda^2$ or a slow logarithmic growth at large $L$ (corresponding to diffusive conduction, see Appendix \ref{appendix1}); we suspect that discriminating between the two cases is not possible with the 
   data we have. 
   If $\lambda^2$ does saturate, this might be ascribed to the suppression of the diffusion constant $D(\epsilon=0) = 0$, which in turn occurs  
    because of the algebraic decay of the average transmission coefficient $T \propto L^{-\gamma}$, $\gamma \approx 0.5$ \cite{Economou}.
 
   It is important to highlight a difference between the Riemann lattice and off-diagonal Anderson model at half-filling: while in the latter $\lambda^2$ appears to saturate in the thermodynamic limit, 
   possibly implying insulating behaviour (consistently with the 
   arguments of Refs~\citen{Economou,Fleishman}), in the former we observe a step-like divergence, characterized by large plateaus and sharp jumps; we conjecture this step-like behaviour to be 
   rooted in certain $-$ as yet unknown $-$ structured correlations present in the nontrivial zeros of $\zeta(s)$.
   
%%%%%%%%%%%%%%%%%%%%%%%%%%%%%%%%%%%%%%%%%%%%%%%%%%%%%%%%
%%%%%%%%%%%%%%%%%%%%%%%%%%%
\vspace{-0.5cm}
\section{Conclusions}
\label{Conclusions}
We investigated the bulk conduction properties of two chirally symmetric aperiodic chains, the Fibonacci and the Riemann lattices, and 
we made comparisons with the off-diagonal Anderson model which features non-deterministic disorder. The chiral symmetry $\tilde{\sigma_z} H \tilde{\sigma_z} = -H$ 
in disordered systems is generally associated to  
anomalies in the single-particle localization length when the Fermi energy is at the band centre \cite{Dyson2}.

We have tested two measures of localization: the participation ratio (PR) and Kohn's many-particle localization length $\lambda$. 
The former is a measure of the spatial extent (extended versus localized) of the single-particle eigenfunctions. The latter, as defined within the modern theory of the insulating state through the kernel 
$P(\textbf{r}, \textbf{r}') = \sum_{j=1}^{N/2}\phi_j(\textbf{r})\phi^{*}_j(\textbf{r}')$, is non-local in energy-space and can therefore capture 
insulating behaviour that arise from a variety of mechanisms \cite{SWM}, not only when the single-particle eigenstate at the Fermi energy is localized.
We have shown this to be particularly important in the case of the Fibonacci lattice, where at specific electron densities -- which are not signalled by the values of PR -- 
insulation occurs due to spectral gaps.

In the Fibonacci lattice, where the one-particle spectrum is singular continuous with a hierarchy of mini-gaps, 
the half and quarter filled systems are found to be metallic; this is likely true for most fillings. 
This indicates that according to the modern theory of the insulating state the Fibonacci quasicrystal is, in general, a metallic system. 
However, at certain specific electronic densities, some of which are given 
by $\rho = 1/g^n$ for integer $n$ and $g$ being the golden ratio, the many-particle system displays insulating behaviour; as anticipated above,
this is seen to be due to the presence of mini-gaps in the single particle spectrum.

The Riemann lattice $-$ 
defined from the location of the nontrivial zeros of the Riemann zeta function $-$ is revealed to possess intriguing bulk insulating 
properties: our results indicate that \text{only} at half-filling will the system display anomalous increase in conduction, while the electron gas is insulating at all other fillings. 
This behaviour is similar to that of the off-diagonal Anderson model, meaning that the zeros of the Riemann zeta function define a model which is more like a non-deterministic random model, 
rather than a quasicrystal, in apparent contrast to Dyson's proposal (see also Ref. \citen{VarmaFuture}).
Still, a difference with respect to the off-diagonal Anderson model emerges: while $\lambda$ displays a step-like divergence at half-filling in the Riemann lattice, possibly indicating 
some hitherto undiscovered long-range correlations present in the Riemann zeta function, a smooth dependence is found in the off-diagonal Anderson model, with a linear increase for small system sizes, 
followed by what appears to be a saturation.\\

We acknowledge interesting discussions with M. Ghulinyan.

   \section*{Appendix}

\subsection{Kohn's localization}
 \label{appendix1}
 \subsubsection*{Kohn's vs. Anderson's localization}
Note that the relation between $\lambda^2$ and conductivity $\sigma(\omega)$ is given by Eq. \eqref{eq: FlucDiss}.
We will illustrate that for Anderson localization with no spectral gaps
\begin{equation}
 \label{eq: Sufficiency}
 \lim_{L \rightarrow \infty} \lambda^2 \neq \infty \Leftrightarrow \sigma(\omega \rightarrow 0) = 0,
\end{equation}
in most physically relevant cases. This absence of conduction as signalled by $\lambda^2$ is referred to as Kohn's localization \cite{Resta, SWM, Bendazolli, Varma}. 
It was shown in Ref. \citen{SWM} that for an insulator $\lambda^2$ saturates and captures the localization of generalised Wannier functions in a higher-dimensional configuration space. 
Therefore $\lambda^2$ will reflect the spectral properties (such as density of states) as well, whereas Anderson localization deals with only 
single-particle eigenstates. In this section we will explicate the connection between the two types of localization.

Let us consider various scenarios of single-particle eigenstate localization. First let us note that the only situation where Eq. \eqref{eq: Sufficiency} is not satisfied will be 
\begin{equation}
 \sigma(\omega) \propto 1/[\log{(1/\omega)}]^{b},      \hspace{3em}     \textrm{with}\hspace{1em} 0<b\leq 1.
\end{equation}

(i) \textit{Power-law localization}: With $|\phi|^2 \propto R^{-\mu} = \exp{(-\mu\log{R})}$, then resonant pairs separated by small energy $\omega$ are found at sites separated by distance $r$ such that
$\omega \sim W\exp{(-\mu\log{r})}$,  giving
\begin{equation}
 r \sim (W/\omega)^{1/\mu}
\end{equation}
for some microscopic energy scale $W$. 

Now the current matrix element $j \sim r \omega$ and number of such resonant pairs $\sim r^{d-1}$; then Kubo linear response gives
\begin{equation}
\label{eq: AlgLoc}
\sigma(\omega) \sim (W/\omega)^{(d-1)/\mu} \omega ^{2 - 2/\mu} \sim \omega ^ {2 - (d+1)/\mu}.  
\end{equation}
Substituting Eq. \eqref{eq: AlgLoc} into Eq. \eqref{eq: FlucDiss} we see that Eq. \eqref{eq: Sufficiency} is satisfied.

(ii) \textit{Exponential localization}: Consider the case $\mu = \infty$ e.g. a (stretched) exponential, with $|\phi|^2 = \exp{(-b R^{\alpha})}$, and $\alpha \neq 1$. 
In this case same arguments go through and we get
\begin{equation}
 \sigma(\omega) \sim \omega^2 \log^{(d+1)/\alpha}{(\omega)}.
\end{equation}
Note that for $\alpha = 1$ we recover the usual Mott-conductivity. In these cases too Eq. \eqref{eq: Sufficiency} is satisfied.\\

(iii) \textit{Power-log localization}: For the hypothetical case $|\phi|^{2} \propto R^{-1} (\log{R})^{-g}$, we may approximate the low-frequency conductivity as
\begin{equation}
\sigma(\omega) \propto [\log{(1/\omega)}]^{-2g},
\end{equation}
which goes to zero as $\omega \rightarrow 0$. The dominant contribution to $\lambda^2 \sim \frac{\sigma(\omega)}{\omega}$ diverges only if $g<1/2$. However $g>1$ is required for
normalization of $|\phi|^2$. 
Hence, here too, for power-log localization Eq. \eqref{eq: Sufficiency} is satisfied i.e. Anderson localization implies Kohn's localization.

\subsubsection*{Scaling of $\lambda^2$}

We will now investigate how $\lambda^2$ is expected to scale with system size $L$ for three possible regimes of transport in the many-particle system: 

(i) \textit{Insulating regime}: here $\sigma(\omega \rightarrow 0) = \omega^{\alpha}$ with some power-law. 
The results of the previous section imply that in this case, for large $L$, $\lambda^2 = \textrm{const.}$. 

(ii) \textit{Diffusive regime}: here $\sigma(\omega \rightarrow 0) = \sigma_0$, a constant for $\omega \in [\Delta, \omega_0]$, where the mean-level spacing $\Delta \propto 1/L$.
This gives for small $\omega$ (or large $L$)
\begin{equation}
 \lambda^2 \sim \sigma_0 \log{\frac{\omega_0}{\Delta}} \sim \log{L},
\end{equation}
in the diffusive or conductive regime.

(iii) \textit{Ballistic regime}: here $\sigma(\omega \rightarrow 0) \propto L$. This can be seen by the following analysis. 

A steady state current $j$ is defined through a diffusion equation as
\begin{equation}
\label{eq: current}
j = D(L) (n(1) - n(L))/L
\end{equation}
with a length-dependent diffusion constant $D(L)$; $n(i)$ is the particle-number at site $i$.
Diffusive transport corresponds to $D(L)=\textrm{const.}$. Then $j \sim 1/L$ at a fixed particle density difference. Then by definition $j \sim 1/L^{x}$ generically. $x>1$ is subdiffusion, 
$x<1$ is superdiffusion. A limiting $x=0$ corresponds to the ballistic transport. 
From the definition $j\sim L^{-x}$ and the generalized diffusion Eq.\eqref{eq: current} it follows that $D(L) \sim L^{1-x}$ [in particular $D(L) \sim L$ for ballistic transport].
Using this and the Einstein relation Eq. \eqref{eq: Einstein} $\sigma(\omega) \propto D(\omega)$, we arrive at
\begin{equation}
 \label{eq: conductivityOmega0}
 \sigma(\omega \rightarrow 0) \propto L^{1-x}.
\end{equation}
This shows that \textit{only} for $x=0$ i.e. ballistic transport, will
\begin{equation}
\sigma(\omega \rightarrow 0) \propto L 
\end{equation}
be valid. An additional factor of $\log(L)$ to $\lambda^2$ should come from the Thouless contribution as for the diffusive case.

Now let us consider the Drude peak that must appear in the ballistic regime. We will show now that this also gives a contribution $\propto L$  in the ballistic regime. 
Indeed, the $\delta_D$-function in  $\int (\textrm{d}\omega/\omega) \delta_D(\omega)$ should be broadened with the width $\Gamma \propto 1/L$. 
The continuous approximation works as long as $\omega >\Delta$, the mean level spacing. So, the integral
\begin{eqnarray}
 \int \frac{\textrm{d}\omega}{\omega} \delta_D(\omega) &=& \int_{\Delta} \frac{\textrm{d}\omega}{\omega} \frac{\Gamma}{\pi (\omega^{2}+\Gamma^{2})} \nonumber \\
 &=& \frac{1}{\pi \Gamma} \int_{\Delta}^{\Gamma} \frac{\textrm{d}\omega}{\omega} \sim L \int_{\Delta}^{\Gamma} \frac{\textrm{d}\omega}{\omega}.
\end{eqnarray}
The remaining integral can give at most a contribution $\log{L}$. So, the Drude peak contributes similarly as the regular part to $\lambda^2$ in the ballistic regime. 
   
\subsection{Fibonacci quasicrystal}
 \label{appendix2} 
Consider a system constructed from the two symbols ${L,S}$ by the substitution rule 
\begin{equation}
 \label{eq: FibDefn}
 \left( 
 \begin{array}{c} 
 L \\ S 
 \end{array} 
 \right) 
 \rightarrow 
 \left(
 \begin{array}{cc} 
 1 & 1 \\ 1 & 0 
 \end{array} \right)
  \left( \begin{array}{c} L \\ S \end{array} \right).
\end{equation}
The transformation matrix has the eigenvalues $g, 1/g$ where $g$ is the golden ratio. 
Repeated application of the above rule gives the sequence of the \textit{Fibonacci quasicrystal} 
\begin{eqnarray}
 && L \nonumber \\
 && LS \nonumber \\
 && LSL \nonumber \\
 && LSLLS \nonumber \\
 && LSLLSLSL \nonumber \\
 && LSLLSLSLLSLLS \nonumber \\
 && \vdots
\end{eqnarray}

The two symbols can define bond-lengths whose ratio $L/S$ is generally taken to be $g$ \cite{Steinhardt, Senechal}, 
an irrational number. This latter condition makes the two underlying lattices incommensurate with one another, and thence giving non-overlapping Fourier components.
\begin{figure}[ttp]
  \centering  
    \begin{tikzpicture}
    \coordinate (Origin)   at (0,0);
    \coordinate (XAxisMin) at (0,0);
    \coordinate (XAxisMax) at (5,0);
    \coordinate (YAxisMin) at (0,0);
    \coordinate (YAxisMax) at (0,3);
    \draw [thick, black,-latex] (XAxisMin) -- (XAxisMax);% Draw x axis
    \draw [thick, black,-latex] (YAxisMin) -- (YAxisMax);% Draw y axis
    \draw [thick, red] (0,0) -- (6*1,6/1.6180);
      \draw [thick, red] (0,0) -- (-2*1,-2/1.6180);
    \clip (-2,-1) rectangle (6.5cm,3.5cm); % Clips the picture...

    \coordinate (Bone) at (0,2);
    \coordinate (Btwo) at (2,-2);
    
    \draw[style=help lines,gray, thin, dashed] (-4,-2) grid[step=1cm] (6,6);

    \foreach \x in {-16,-14,...,20}{% Two, indices running over each
      \foreach \y in {-12,-10,...,12}{% node on the grid we have drawn 
        \node[draw,circle,inner sep=1pt,fill] at (0.5*\x,0.5*\y) {};
            % Places a dot at those points
      }
    }
    
     \draw [thick, blue] (-1,-1) -- (-1.1708,-0.7236);
     \draw [thick, blue] (-1,0) -- (-0.7236,-0.4472);
     \draw [thick, blue] (1,0) -- (0.72361,0.44721);
     \draw [thick, blue] (1,1) -- (1.17082,0.72361);
     \draw [thick, blue] (2,1) -- (1.89443,1.17082);
     \draw [thick, blue] (2,2) -- (2.3416,1.4472);
     \draw [thick, blue] (3,2) -- (3.0652,1.8944);
     \draw [thick, blue] (4,2) -- (3.7888,2.3416);
     \draw [thick, blue] (4,3) -- (4.236,2.618);
     \draw [thick, blue] (5,3) -- (4.9596,3.0652);
     \draw [thick, blue] (6,3) -- (5.6832,3.5214);
    
  \end{tikzpicture}
  
  \caption{Cut and project method for constructing the Fibonacci quasicrystal. The full red line has slope $1/g$ and the projections of the underlying square lattice onto this line 
  generates the Fibonacci quasicrystal. The lattice distances along this line are given by Eq. \eqref{eq: FibSeq}.}
  \label{fig: CP}
\end{figure}
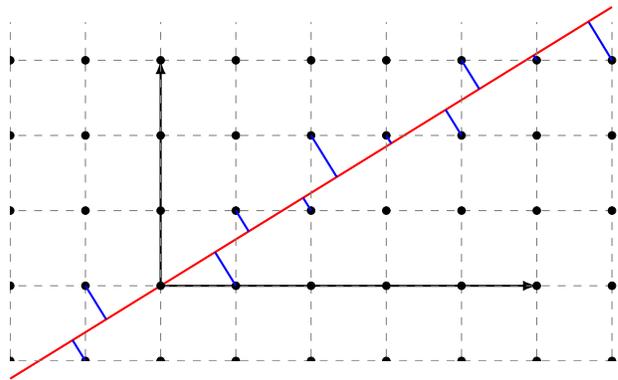
The Fourier transform of the Fibonacci lattice is defined as usual by 
\begin{equation}
 \label{eq: Fourier}
F(k) = \mathcal{F}\left\{\sum_{n}\delta(x-x_n)\right\}.
\end{equation}
Now the lattice positions $x_n$ may be computed from a Beatty sequence (lower Wythoff sequence) that indexes the Fibonacci word \cite{Wythoff} or by a cut and project method \cite{Senechal} from 
a 2D square lattice
\begin{equation}
 \label{eq: FibSeq}
 x_n = n + (g-1)E[(n+1)/g],
 \end{equation}
with $E$ being the floor function. In the absence of the second term, the lattice is a usual periodic lattice with $x_n = n$. The construction using the cut and project method \cite{Senechal} is 
shown in Fig. \ref{fig: CP}.
Using the above, Eq. \eqref{eq: Fourier} is simplified to
\begin{equation}
 \label {eq: FibFourier}
 F(k) = \sum_{l,m}F_{l,m}\delta(k - k_{l,m}).
\end{equation}
Here $k_{l,m}$ is given by \cite{Steinhardt}
\begin{equation}
 k_{l,m} = \frac{2\pi g^2}{1+g^2} (l + m/g)
\end{equation}
for integers $l,m$. 

Eqs. \eqref{eq: FibSeq} and \eqref{eq: FibFourier} show that the diffraction pattern of the Fibonacci quasicrystal is well-defined and densely fills the reciprocal space due to the 
incommensurability of the two underlying lattices in Eq. \eqref{eq: FibSeq} (thereby giving two summations instead of one in Eq. \eqref{eq: FibFourier}).

\subsection{Multifractal analysis}
\label{appendix3}
The multifractal analysis of an eigenstate $\phi$ with given energy $E$ generalizes the inverse PR to all $q$ moments of the wavefunction amplitudes, 
for $q \in [-\infty, \infty]$, and is based on the usual box-counting procedure \cite{Chhabra, Schreiber, Grussbach}. 
The probability measure $\mu_k(\delta)$ of finding a particle in the $k^{\textrm{th}}$ box of linear size $a \ll l \ll L$, 
where $a$ is some averaged lattice spacing, such that $\delta \equiv l/L$, is given by
\begin{equation}
 \label{eq: probmeasure}
 \mu_k(\delta) = \sum_{i_k}|\phi({i_k})|^2,
\end{equation}
where $i_k$ are the site indices in the $k^{\textrm{th}}$ box. The $q^{\textrm{th}}$ moment of the probabality measure $\mu_k(\delta)$ is then defined as
\begin{equation}
 \label{eq: qmeasure}
 \mu_k(q,\delta) = \frac{\mu^q_k(\delta)}{\sum_k\mu^q_k(\delta)},
\end{equation}
with the $k$-summation being over all boxes. With this the Hausdorff dimension $f$ that measures the multifractality of the eigenstate under consideration is given parametrically 
in terms of the moments $q$ as
\begin{equation}
 \label{eq: falpha}
 f(q) = \lim_{\delta\rightarrow 0}\sum_k\mu_k(q,\delta)\ln{(\mu_k(q,\delta))}/\ln{(\delta)},
\end{equation}
with the Lipshitz-H\"{o}lder index $\alpha$ given parametrically as 
\begin{equation}
 \label{eq: alpha}
 \alpha(q) = \lim_{\delta\rightarrow 0}\sum_k\mu_k(q,\delta)\ln{(\mu_k(1,\delta))}/\ln{(\delta)}.
\end{equation}
The Hausdorff dimension $f(\alpha(q))$ measures the fraction of boxes $L^{f(\alpha)}$ that scale as $\alpha$ i.e. $L^{-\alpha}$; 
in the rest we choose a range of box sizes $l = [l_1,l_2]$, 
and perform a linear least-squares fit of the numerators in \eqref{eq: falpha}, \eqref{eq: alpha} to $\ln{(n)}$. 
$q = \infty (-\infty)$ corresponds to the minimum (maximum) 
value of $\alpha$, whereas for $q=0$ the fractal dimension $f(\alpha)$ peaks to the maximum value of $f(\alpha = \alpha_c) = 1$, the \textit{integer} dimension 
of the underlying lattice; for the first moment $q=1$, $f(\alpha) = \alpha$.

\bibliography{Ref}

\begin{thebibliography}{62}
\expandafter\ifx\csname natexlab\endcsname\relax\def\natexlab#1{#1}\fi
\expandafter\ifx\csname bibnamefont\endcsname\relax
  \def\bibnamefont#1{#1}\fi
\expandafter\ifx\csname bibfnamefont\endcsname\relax
  \def\bibfnamefont#1{#1}\fi
\expandafter\ifx\csname citenamefont\endcsname\relax
  \def\citenamefont#1{#1}\fi
\expandafter\ifx\csname url\endcsname\relax
  \def\url#1{\texttt{#1}}\fi
\expandafter\ifx\csname urlprefix\endcsname\relax\def\urlprefix{URL }\fi
\providecommand{\bibinfo}[2]{#2}
\providecommand{\eprint}[2][]{\url{#2}}

\bibitem[{\citenamefont{Shechtman et~al.}(1984)\citenamefont{Shechtman, Blech,
  Gratias, and Cahn}}]{Shechtman}
\bibinfo{author}{\bibfnamefont{D.}~\bibnamefont{Shechtman}},
  \bibinfo{author}{\bibfnamefont{I.}~\bibnamefont{Blech}},
  \bibinfo{author}{\bibfnamefont{D.}~\bibnamefont{Gratias}}, \bibnamefont{and}
  \bibinfo{author}{\bibfnamefont{J.~W.} \bibnamefont{Cahn}},
  \bibinfo{journal}{Phys. Rev. Lett.} \textbf{\bibinfo{volume}{53}},
  \bibinfo{pages}{1951} (\bibinfo{year}{1984}).

\bibitem[{\citenamefont{Levine and Steinhardt}(1984)}]{Steinhardt}
\bibinfo{author}{\bibfnamefont{D.}~\bibnamefont{Levine}} \bibnamefont{and}
  \bibinfo{author}{\bibfnamefont{P.~J.} \bibnamefont{Steinhardt}},
  \bibinfo{journal}{Phys. Rev. Lett.} \textbf{\bibinfo{volume}{53}},
  \bibinfo{pages}{2477} (\bibinfo{year}{1984}).

\bibitem[{\citenamefont{Senechal}(1996)}]{Senechal}
\bibinfo{author}{\bibfnamefont{M.}~\bibnamefont{Senechal}},
  \emph{\bibinfo{title}{Quasicrystals and geometry}}
  (\bibinfo{publisher}{Cambridge University Press}, \bibinfo{year}{1996}).

\bibitem[{\citenamefont{Hof}(1995)}]{Hof}
\bibinfo{author}{\bibfnamefont{A.}~\bibnamefont{Hof}}, \bibinfo{journal}{Comm.
  Math. Phys.} \textbf{\bibinfo{volume}{169}}, \bibinfo{pages}{25}
  (\bibinfo{year}{1995}).

\bibitem[{Not()}]{Notediffraction}
\bibinfo{note}{Technically, the diffraction pattern corresponds to the Fourier
  transform of the density-density correlations.}

\bibitem[{\citenamefont{Aubry et~al.}(1995)\citenamefont{Aubry, Godr\'eche, and
  Luck}}]{AubryLuck}
\bibinfo{author}{\bibfnamefont{S.}~\bibnamefont{Aubry}},
  \bibinfo{author}{\bibfnamefont{C.}~\bibnamefont{Godr\'eche}},
  \bibnamefont{and} \bibinfo{author}{\bibfnamefont{J.~M.} \bibnamefont{Luck}},
  \bibinfo{journal}{Comm. Math. Phys.} \textbf{\bibinfo{volume}{169}},
  \bibinfo{pages}{25} (\bibinfo{year}{1995}).

\bibitem[{\citenamefont{Merlin et~al.}(1985)\citenamefont{Merlin, Bajema,
  Clarke, Juang, and Bhattacharya}}]{MerlinExpt}
\bibinfo{author}{\bibfnamefont{R.}~\bibnamefont{Merlin}},
  \bibinfo{author}{\bibfnamefont{K.}~\bibnamefont{Bajema}},
  \bibinfo{author}{\bibfnamefont{R.}~\bibnamefont{Clarke}},
  \bibinfo{author}{\bibfnamefont{F.-Y.} \bibnamefont{Juang}}, \bibnamefont{and}
  \bibinfo{author}{\bibfnamefont{P.~K.} \bibnamefont{Bhattacharya}},
  \bibinfo{journal}{Phys. Rev. Lett.} \textbf{\bibinfo{volume}{55}},
  \bibinfo{pages}{1768} (\bibinfo{year}{1985}).

\bibitem[{\citenamefont{Dharma-wardana
  et~al.}(1987)\citenamefont{Dharma-wardana, MacDonald, Lockwood, Baribeau, and
  Houghton}}]{Houghton}
\bibinfo{author}{\bibfnamefont{M.~W.~C.} \bibnamefont{Dharma-wardana}},
  \bibinfo{author}{\bibfnamefont{A.~H.} \bibnamefont{MacDonald}},
  \bibinfo{author}{\bibfnamefont{D.~J.} \bibnamefont{Lockwood}},
  \bibinfo{author}{\bibfnamefont{J.-M.} \bibnamefont{Baribeau}},
  \bibnamefont{and} \bibinfo{author}{\bibfnamefont{D.~C.}
  \bibnamefont{Houghton}}, \bibinfo{journal}{Phys. Rev. Lett.}
  \textbf{\bibinfo{volume}{58}}, \bibinfo{pages}{1761} (\bibinfo{year}{1987}).

\bibitem[{\citenamefont{Thiel and Dubois}(2000)}]{Dubois}
\bibinfo{author}{\bibfnamefont{P.~A.} \bibnamefont{Thiel}} \bibnamefont{and}
  \bibinfo{author}{\bibfnamefont{J.~M.} \bibnamefont{Dubois}},
  \bibinfo{journal}{Nature: News and Views} \textbf{\bibinfo{volume}{406}},
  \bibinfo{pages}{570} (\bibinfo{year}{2000}).

\bibitem[{\citenamefont{Martin et~al.}(1991)\citenamefont{Martin, Hebard,
  Kortan, and Thiel}}]{Martin}
\bibinfo{author}{\bibfnamefont{S.}~\bibnamefont{Martin}},
  \bibinfo{author}{\bibfnamefont{A.~F.} \bibnamefont{Hebard}},
  \bibinfo{author}{\bibfnamefont{A.~R.} \bibnamefont{Kortan}},
  \bibnamefont{and} \bibinfo{author}{\bibfnamefont{F.~A.} \bibnamefont{Thiel}},
  \bibinfo{journal}{Phys. Rev. Lett.} \textbf{\bibinfo{volume}{67}},
  \bibinfo{pages}{719} (\bibinfo{year}{1991}).

\bibitem[{\citenamefont{Roati et~al.}(2008)\citenamefont{Roati, D'Errico,
  Fallani, Fattori, Fort, Zaccanti, Modugno, Modugno, and Inguscio}}]{Roati}
\bibinfo{author}{\bibfnamefont{G.}~\bibnamefont{Roati}},
  \bibinfo{author}{\bibfnamefont{C.}~\bibnamefont{D'Errico}},
  \bibinfo{author}{\bibfnamefont{L.}~\bibnamefont{Fallani}},
  \bibinfo{author}{\bibfnamefont{M.}~\bibnamefont{Fattori}},
  \bibinfo{author}{\bibfnamefont{C.}~\bibnamefont{Fort}},
  \bibinfo{author}{\bibfnamefont{M.}~\bibnamefont{Zaccanti}},
  \bibinfo{author}{\bibfnamefont{G.}~\bibnamefont{Modugno}},
  \bibinfo{author}{\bibfnamefont{M.}~\bibnamefont{Modugno}}, \bibnamefont{and}
  \bibinfo{author}{\bibfnamefont{M.}~\bibnamefont{Inguscio}},
  \bibinfo{journal}{Nature} \textbf{\bibinfo{volume}{453}},
  \bibinfo{pages}{895} (\bibinfo{year}{2008}).

\bibitem[{\citenamefont{Schreiber et~al.}(2015)\citenamefont{Schreiber,
  Hodgman, Bordia, L�schen, Fischer, Vosk, Altman, Schneider, and
  Bloch}}]{Bloch}
\bibinfo{author}{\bibfnamefont{M.}~\bibnamefont{Schreiber}},
  \bibinfo{author}{\bibfnamefont{S.~S.} \bibnamefont{Hodgman}},
  \bibinfo{author}{\bibfnamefont{P.}~\bibnamefont{Bordia}},
  \bibinfo{author}{\bibfnamefont{H.~P.} \bibnamefont{L�schen}},
  \bibinfo{author}{\bibfnamefont{M.~H.} \bibnamefont{Fischer}},
  \bibinfo{author}{\bibfnamefont{R.}~\bibnamefont{Vosk}},
  \bibinfo{author}{\bibfnamefont{E.}~\bibnamefont{Altman}},
  \bibinfo{author}{\bibfnamefont{U.}~\bibnamefont{Schneider}},
  \bibnamefont{and} \bibinfo{author}{\bibfnamefont{I.}~\bibnamefont{Bloch}},
  \bibinfo{journal}{Science} \textbf{\bibinfo{volume}{349}},
  \bibinfo{pages}{842} (\bibinfo{year}{2015}).

\bibitem[{\citenamefont{Singh et~al.}(2015)\citenamefont{Singh, Saha,
  Parameswaran, and Weld}}]{Singh}
\bibinfo{author}{\bibfnamefont{K.}~\bibnamefont{Singh}},
  \bibinfo{author}{\bibfnamefont{K.}~\bibnamefont{Saha}},
  \bibinfo{author}{\bibfnamefont{S.~A.} \bibnamefont{Parameswaran}},
  \bibnamefont{and} \bibinfo{author}{\bibfnamefont{D.~M.} \bibnamefont{Weld}},
  \bibinfo{journal}{Phys. Rev. A} \textbf{\bibinfo{volume}{92}},
  \bibinfo{pages}{063426} (\bibinfo{year}{2015}).

\bibitem[{\citenamefont{Lahini et~al.}(2009)\citenamefont{Lahini, Pugatch,
  Pozzi, Sorel, Morandotti, Davidson, and Silberberg}}]{Lahini}
\bibinfo{author}{\bibfnamefont{Y.}~\bibnamefont{Lahini}},
  \bibinfo{author}{\bibfnamefont{R.}~\bibnamefont{Pugatch}},
  \bibinfo{author}{\bibfnamefont{F.}~\bibnamefont{Pozzi}},
  \bibinfo{author}{\bibfnamefont{M.}~\bibnamefont{Sorel}},
  \bibinfo{author}{\bibfnamefont{R.}~\bibnamefont{Morandotti}},
  \bibinfo{author}{\bibfnamefont{N.}~\bibnamefont{Davidson}}, \bibnamefont{and}
  \bibinfo{author}{\bibfnamefont{Y.}~\bibnamefont{Silberberg}},
  \bibinfo{journal}{Phys. Rev. Lett.} \textbf{\bibinfo{volume}{103}},
  \bibinfo{pages}{013901} (\bibinfo{year}{2009}).

\bibitem[{\citenamefont{Ghulinyan}(2015)}]{Mher}
\bibinfo{author}{\bibfnamefont{M.}~\bibnamefont{Ghulinyan}}, in
  \emph{\bibinfo{booktitle}{Light Localisation and Lasing}}, edited by
  \bibinfo{editor}{\bibfnamefont{M.}~\bibnamefont{Ghulinyan}} \bibnamefont{and}
  \bibinfo{editor}{\bibfnamefont{L.}~\bibnamefont{Pavesi}}
  (\bibinfo{publisher}{Cambridge Univ. Press}, \bibinfo{year}{2015}),
  chap.~\bibinfo{chapter}{5}.

\bibitem[{\citenamefont{Kohmoto et~al.}(1983)\citenamefont{Kohmoto, Kadanoff,
  and Tang}}]{Kadanoff}
\bibinfo{author}{\bibfnamefont{M.}~\bibnamefont{Kohmoto}},
  \bibinfo{author}{\bibfnamefont{L.~P.} \bibnamefont{Kadanoff}},
  \bibnamefont{and} \bibinfo{author}{\bibfnamefont{C.}~\bibnamefont{Tang}},
  \bibinfo{journal}{Phys. Rev. Lett.} \textbf{\bibinfo{volume}{50}},
  \bibinfo{pages}{1870} (\bibinfo{year}{1983}).

\bibitem[{\citenamefont{Ostlund et~al.}(1983)\citenamefont{Ostlund, Pandit,
  Rand, Schellnhuber, and Siggia}}]{Siggia}
\bibinfo{author}{\bibfnamefont{S.}~\bibnamefont{Ostlund}},
  \bibinfo{author}{\bibfnamefont{R.}~\bibnamefont{Pandit}},
  \bibinfo{author}{\bibfnamefont{D.}~\bibnamefont{Rand}},
  \bibinfo{author}{\bibfnamefont{H.~J.} \bibnamefont{Schellnhuber}},
  \bibnamefont{and} \bibinfo{author}{\bibfnamefont{E.}~\bibnamefont{Siggia}},
  \bibinfo{journal}{Phys. Rev. Lett.} \textbf{\bibinfo{volume}{50}},
  \bibinfo{pages}{1873} (\bibinfo{year}{1983}).

\bibitem[{\citenamefont{Fujiwara et~al.}(1989)\citenamefont{Fujiwara, Kohmoto,
  and Tokihiro}}]{Fujiwara}
\bibinfo{author}{\bibfnamefont{T.}~\bibnamefont{Fujiwara}},
  \bibinfo{author}{\bibfnamefont{M.}~\bibnamefont{Kohmoto}}, \bibnamefont{and}
  \bibinfo{author}{\bibfnamefont{T.}~\bibnamefont{Tokihiro}},
  \bibinfo{journal}{Phys. Rev. B(R)} \textbf{\bibinfo{volume}{40}},
  \bibinfo{pages}{7413} (\bibinfo{year}{1989}).

\bibitem[{\citenamefont{Rotenberg et~al.}(2000)\citenamefont{Rotenberg, Theis,
  Horn, and Gille}}]{Rotenberg}
\bibinfo{author}{\bibfnamefont{E.}~\bibnamefont{Rotenberg}},
  \bibinfo{author}{\bibfnamefont{W.}~\bibnamefont{Theis}},
  \bibinfo{author}{\bibfnamefont{K.}~\bibnamefont{Horn}}, \bibnamefont{and}
  \bibinfo{author}{\bibfnamefont{P.}~\bibnamefont{Gille}},
  \bibinfo{journal}{Nature} \textbf{\bibinfo{volume}{406}},
  \bibinfo{pages}{602} (\bibinfo{year}{2000}).

\bibitem[{\citenamefont{Anderson}(1958)}]{Anderson}
\bibinfo{author}{\bibfnamefont{P.~W.} \bibnamefont{Anderson}},
  \bibinfo{journal}{Phys. Rev.} \textbf{\bibinfo{volume}{109}},
  \bibinfo{pages}{1492} (\bibinfo{year}{1958}).

\bibitem[{\citenamefont{Abrahams et~al.}(1979)\citenamefont{Abrahams, Anderson,
  Licciardello, and Ramakrishnan}}]{Gang4}
\bibinfo{author}{\bibfnamefont{E.}~\bibnamefont{Abrahams}},
  \bibinfo{author}{\bibfnamefont{P.~W.} \bibnamefont{Anderson}},
  \bibinfo{author}{\bibfnamefont{D.~C.} \bibnamefont{Licciardello}},
  \bibnamefont{and} \bibinfo{author}{\bibfnamefont{T.~V.}
  \bibnamefont{Ramakrishnan}}, \bibinfo{journal}{Phys. Rev. Lett.}
  \textbf{\bibinfo{volume}{42}}, \bibinfo{pages}{673} (\bibinfo{year}{1979}).

\bibitem[{\citenamefont{Kohn}(1963)}]{Kohn}
\bibinfo{author}{\bibfnamefont{W.}~\bibnamefont{Kohn}}, \bibinfo{journal}{Phys.
  Rev.} \textbf{\bibinfo{volume}{133}}, \bibinfo{pages}{A171}
  (\bibinfo{year}{1963}).

\bibitem[{\citenamefont{Resta and Sorella}(1999)}]{RS}
\bibinfo{author}{\bibfnamefont{R.}~\bibnamefont{Resta}} \bibnamefont{and}
  \bibinfo{author}{\bibfnamefont{S.}~\bibnamefont{Sorella}},
  \bibinfo{journal}{Phys. Rev. Lett.} \textbf{\bibinfo{volume}{82}},
  \bibinfo{pages}{370} (\bibinfo{year}{1999}).

\bibitem[{\citenamefont{Souza et~al.}(2000)\citenamefont{Souza, Wilkens, and
  Martin}}]{SWM}
\bibinfo{author}{\bibfnamefont{I.}~\bibnamefont{Souza}},
  \bibinfo{author}{\bibfnamefont{T.}~\bibnamefont{Wilkens}}, \bibnamefont{and}
  \bibinfo{author}{\bibfnamefont{R.~M.} \bibnamefont{Martin}},
  \bibinfo{journal}{Phys. Rev. B} \textbf{\bibinfo{volume}{62}},
  \bibinfo{pages}{1666} (\bibinfo{year}{2000}).

\bibitem[{\citenamefont{Resta}(2002)}]{RestaTopical}
\bibinfo{author}{\bibfnamefont{R.}~\bibnamefont{Resta}}, \bibinfo{journal}{J.
  Phys.: Condens. Matter} \textbf{\bibinfo{volume}{14}}, \bibinfo{pages}{R625}
  (\bibinfo{year}{2002}).

\bibitem[{\citenamefont{Resta}(2011)}]{Resta}
\bibinfo{author}{\bibfnamefont{R.}~\bibnamefont{Resta}}, \bibinfo{journal}{Eur.
  Phys. J. B} \textbf{\bibinfo{volume}{79}}, \bibinfo{pages}{121}
  (\bibinfo{year}{2011}).

\bibitem[{\citenamefont{Bendazolli et~al.}(2010)\citenamefont{Bendazolli,
  Evangelisti, Monari, and Resta}}]{Bendazolli}
\bibinfo{author}{\bibfnamefont{G.~L.} \bibnamefont{Bendazolli}},
  \bibinfo{author}{\bibfnamefont{S.}~\bibnamefont{Evangelisti}},
  \bibinfo{author}{\bibfnamefont{A.}~\bibnamefont{Monari}}, \bibnamefont{and}
  \bibinfo{author}{\bibfnamefont{R.}~\bibnamefont{Resta}}, \bibinfo{journal}{J.
  Chem. Phys.} \textbf{\bibinfo{volume}{133}}, \bibinfo{pages}{064703}
  (\bibinfo{year}{2010}).

\bibitem[{\citenamefont{Varma and Pilati}(2015)}]{Varma}
\bibinfo{author}{\bibfnamefont{V.~K.} \bibnamefont{Varma}} \bibnamefont{and}
  \bibinfo{author}{\bibfnamefont{S.}~\bibnamefont{Pilati}},
  \bibinfo{journal}{Phys. Rev. B} \textbf{\bibinfo{volume}{92}},
  \bibinfo{pages}{134207} (\bibinfo{year}{2015}).

\bibitem[{\citenamefont{Dyson}(2009)}]{Dyson}
\bibinfo{author}{\bibfnamefont{F.}~\bibnamefont{Dyson}},
  \emph{\bibinfo{title}{Frogs and Birds}} (\bibinfo{publisher}{Notices of the
  American Mathematical Society}, \bibinfo{year}{2009}).

\bibitem[{Bou()}]{BoundNote}
\bibinfo{note}{By an 'open chan' we do not mean it to be connected to a lead or
  bath but simply that the two ends are not connected to each other by the
  Hamiltonian as in a periodic system.}

\bibitem[{\citenamefont{Kravtsov and Yudson}(2011)}]{Yudson}
\bibinfo{author}{\bibfnamefont{V.~E.} \bibnamefont{Kravtsov}} \bibnamefont{and}
  \bibinfo{author}{\bibfnamefont{V.~I.} \bibnamefont{Yudson}},
  \bibinfo{journal}{Ann. Phys.-New York} \textbf{\bibinfo{volume}{326}},
  \bibinfo{pages}{1672} (\bibinfo{year}{2011}).

\bibitem[{\citenamefont{Johri and Bhatt}(2012)}]{Johri}
\bibinfo{author}{\bibfnamefont{S.}~\bibnamefont{Johri}} \bibnamefont{and}
  \bibinfo{author}{\bibfnamefont{R.~N.} \bibnamefont{Bhatt}},
  \bibinfo{journal}{Phys. Rev. Lett.} \textbf{\bibinfo{volume}{109}},
  \bibinfo{pages}{076402} (\bibinfo{year}{2012}).

\bibitem[{\citenamefont{Olsen et~al.}(2016)\citenamefont{Olsen, Resta, and
  Souza}}]{RestaNEW}
\bibinfo{author}{\bibfnamefont{T.}~\bibnamefont{Olsen}},
  \bibinfo{author}{\bibfnamefont{R.}~\bibnamefont{Resta}}, \bibnamefont{and}
  \bibinfo{author}{\bibfnamefont{I.}~\bibnamefont{Souza}},
  \bibinfo{journal}{arXiv preprint arXiv:1604.01006}  (\bibinfo{year}{2016}).

\bibitem[{\citenamefont{Resta}(2005)}]{RestaQHP}
\bibinfo{author}{\bibfnamefont{R.}~\bibnamefont{Resta}},
  \bibinfo{journal}{Phys. Rev. Lett.} \textbf{\bibinfo{volume}{95}},
  \bibinfo{pages}{196805} (\bibinfo{year}{2005}).

\bibitem[{\citenamefont{Kohmoto et~al.}(1987)\citenamefont{Kohmoto, Sutherland,
  and Tang}}]{Sutherland}
\bibinfo{author}{\bibfnamefont{M.}~\bibnamefont{Kohmoto}},
  \bibinfo{author}{\bibfnamefont{B.}~\bibnamefont{Sutherland}},
  \bibnamefont{and} \bibinfo{author}{\bibfnamefont{C.}~\bibnamefont{Tang}},
  \bibinfo{journal}{Phys. Rev. B} \textbf{\bibinfo{volume}{35}},
  \bibinfo{pages}{1020} (\bibinfo{year}{1987}).

\bibitem[{\citenamefont{Kohmoto and Banavar}(1986)}]{Banavar}
\bibinfo{author}{\bibfnamefont{M.}~\bibnamefont{Kohmoto}} \bibnamefont{and}
  \bibinfo{author}{\bibfnamefont{J.~R.} \bibnamefont{Banavar}},
  \bibinfo{journal}{Phys. Rev. B} \textbf{\bibinfo{volume}{34}},
  \bibinfo{pages}{563} (\bibinfo{year}{1986}).

\bibitem[{\citenamefont{Hiramoto and Kohmoto}(1992)}]{Hiramoto}
\bibinfo{author}{\bibfnamefont{H.}~\bibnamefont{Hiramoto}} \bibnamefont{and}
  \bibinfo{author}{\bibfnamefont{M.}~\bibnamefont{Kohmoto}},
  \bibinfo{journal}{Int. J. Mod. Phys. B} \textbf{\bibinfo{volume}{6}},
  \bibinfo{pages}{281} (\bibinfo{year}{1992}).

\bibitem[{\citenamefont{Chhabra and Jensen}(1989)}]{Chhabra}
\bibinfo{author}{\bibfnamefont{A.}~\bibnamefont{Chhabra}} \bibnamefont{and}
  \bibinfo{author}{\bibfnamefont{R.~V.} \bibnamefont{Jensen}},
  \bibinfo{journal}{Phys. Rev. Lett.} \textbf{\bibinfo{volume}{62}},
  \bibinfo{pages}{1327} (\bibinfo{year}{1989}).

\bibitem[{\citenamefont{Schreiber and Grussbach}(1991)}]{Schreiber}
\bibinfo{author}{\bibfnamefont{M.}~\bibnamefont{Schreiber}} \bibnamefont{and}
  \bibinfo{author}{\bibfnamefont{H.}~\bibnamefont{Grussbach}},
  \bibinfo{journal}{Phys. Rev. Lett.} \textbf{\bibinfo{volume}{67}},
  \bibinfo{pages}{607} (\bibinfo{year}{1991}).

\bibitem[{\citenamefont{Grussbach and Schreiber}(1993)}]{Grussbach}
\bibinfo{author}{\bibfnamefont{H.}~\bibnamefont{Grussbach}} \bibnamefont{and}
  \bibinfo{author}{\bibfnamefont{M.}~\bibnamefont{Schreiber}},
  \bibinfo{journal}{Chem. Phys.} \textbf{\bibinfo{volume}{177}},
  \bibinfo{pages}{733} (\bibinfo{year}{1993}).

\bibitem[{\citenamefont{Sanderson}(2010)}]{Armadillo}
\bibinfo{author}{\bibfnamefont{C.}~\bibnamefont{Sanderson}},
  \bibinfo{journal}{Technical Report NICTA}  (\bibinfo{year}{2010}).

\bibitem[{\citenamefont{Resta}(2006)}]{RestaOBC}
\bibinfo{author}{\bibfnamefont{R.}~\bibnamefont{Resta}},
  \bibinfo{journal}{Phys. Rev. Lett.} \textbf{\bibinfo{volume}{96}},
  \bibinfo{pages}{137601} (\bibinfo{year}{2006}).

\bibitem[{\citenamefont{Imry}(2002)}]{Imry}
\bibinfo{author}{\bibfnamefont{Y.}~\bibnamefont{Imry}},
  \emph{\bibinfo{title}{Introduction to mesoscopic physics}}
  (\bibinfo{publisher}{Oxford University Press}, \bibinfo{year}{2002}).

\bibitem[{\citenamefont{Kubo}(1957)}]{Kubo}
\bibinfo{author}{\bibfnamefont{R.}~\bibnamefont{Kubo}}, \bibinfo{journal}{J.
  Phys. Soc. Japan} \textbf{\bibinfo{volume}{12}}, \bibinfo{pages}{570}
  (\bibinfo{year}{1957}).

\bibitem[{\citenamefont{Sutherland and Kohmoto}(1987)}]{SutherlandKohmoto}
\bibinfo{author}{\bibfnamefont{B.}~\bibnamefont{Sutherland}} \bibnamefont{and}
  \bibinfo{author}{\bibfnamefont{M.}~\bibnamefont{Kohmoto}},
  \bibinfo{journal}{Phys. Rev. B} \textbf{\bibinfo{volume}{36}},
  \bibinfo{pages}{5877} (\bibinfo{year}{1987}).

\bibitem[{\citenamefont{Sarma and Xie}(1988)}]{Sarma}
\bibinfo{author}{\bibfnamefont{S.~D.} \bibnamefont{Sarma}} \bibnamefont{and}
  \bibinfo{author}{\bibfnamefont{X.}~\bibnamefont{Xie}},
  \bibinfo{journal}{Phys. Rev. B} \textbf{\bibinfo{volume}{37}},
  \bibinfo{pages}{1097} (\bibinfo{year}{1988}).

\bibitem[{\citenamefont{Maci\'a and Dom\'{\i}nguez-Adame}(1996)}]{Macia}
\bibinfo{author}{\bibfnamefont{E.}~\bibnamefont{Maci\'a}} \bibnamefont{and}
  \bibinfo{author}{\bibfnamefont{F.}~\bibnamefont{Dom\'{\i}nguez-Adame}},
  \bibinfo{journal}{Phys. Rev. Lett.} \textbf{\bibinfo{volume}{76}},
  \bibinfo{pages}{2957} (\bibinfo{year}{1996}).

\bibitem[{\citenamefont{Luck and Nieuwenhuitz}(1988)}]{Luck}
\bibinfo{author}{\bibfnamefont{J.~M.} \bibnamefont{Luck}} \bibnamefont{and}
  \bibinfo{author}{\bibfnamefont{T.~M.} \bibnamefont{Nieuwenhuitz}},
  \bibinfo{journal}{Europhys. Lett.} \textbf{\bibinfo{volume}{2}},
  \bibinfo{pages}{257} (\bibinfo{year}{1988}).

\bibitem[{\citenamefont{Luck and Petritis}(1986)}]{Luck2}
\bibinfo{author}{\bibfnamefont{J.~M.} \bibnamefont{Luck}} \bibnamefont{and}
  \bibinfo{author}{\bibfnamefont{D.}~\bibnamefont{Petritis}},
  \bibinfo{journal}{J. Stat. Phys.} \textbf{\bibinfo{volume}{42}},
  \bibinfo{pages}{259} (\bibinfo{year}{1986}).

\bibitem[{\citenamefont{Riemann}(1859)}]{Riemann}
\bibinfo{author}{\bibfnamefont{B.}~\bibnamefont{Riemann}},
  \bibinfo{journal}{Monatsberichte der Berliner Akademie}
  p.~\bibinfo{pages}{48} (\bibinfo{year}{1859}).

\bibitem[{Cur()}]{Curiosity}
\bibinfo{note}{It is of some curiosity that Riemann made his conjecture after
  checking the first three nontrivial zeros; today it has been verified to the
  first $10^{13}$ zeros as well as a few at much larger heights $ \sim
  10^{24}$.}

\bibitem[{Var()}]{VarmaFuture}
\bibinfo{note}{V. K. Varma et al. (unpublished).}

\bibitem[{\citenamefont{Odlyzko}(1987)}]{Odlyzko}
\bibinfo{author}{\bibfnamefont{A.~M.} \bibnamefont{Odlyzko}},
  \bibinfo{journal}{Math. Comp.} \textbf{\bibinfo{volume}{48}},
  \bibinfo{pages}{273} (\bibinfo{year}{1987}).

\bibitem[{\citenamefont{Dyson}(1953)}]{Dyson2}
\bibinfo{author}{\bibfnamefont{F.~J.} \bibnamefont{Dyson}},
  \bibinfo{journal}{Phys. Rev.} \textbf{\bibinfo{volume}{92}},
  \bibinfo{pages}{1331} (\bibinfo{year}{1953}).

\bibitem[{\citenamefont{Inui et~al.}(1994)\citenamefont{Inui, Trugman, and
  Abrahams}}]{Inui}
\bibinfo{author}{\bibfnamefont{M.}~\bibnamefont{Inui}},
  \bibinfo{author}{\bibfnamefont{S.~A.} \bibnamefont{Trugman}},
  \bibnamefont{and} \bibinfo{author}{\bibfnamefont{E.}~\bibnamefont{Abrahams}},
  \bibinfo{journal}{Phys. Rev. B} \textbf{\bibinfo{volume}{49}},
  \bibinfo{pages}{3190} (\bibinfo{year}{1994}).

\bibitem[{\citenamefont{McKenzie}(1996)}]{McKenzie}
\bibinfo{author}{\bibfnamefont{R.~H.} \bibnamefont{McKenzie}},
  \bibinfo{journal}{Phys. Rev. Lett.} \textbf{\bibinfo{volume}{77}},
  \bibinfo{pages}{4804} (\bibinfo{year}{1996}).

\bibitem[{\citenamefont{Herbert and Jones}(1971)}]{HerbertJones}
\bibinfo{author}{\bibfnamefont{D.~C.} \bibnamefont{Herbert}} \bibnamefont{and}
  \bibinfo{author}{\bibfnamefont{R.}~\bibnamefont{Jones}}, \bibinfo{journal}{J.
  Phys. C: Solid State Phys.} \textbf{\bibinfo{volume}{4}},
  \bibinfo{pages}{1145} (\bibinfo{year}{1971}).

\bibitem[{\citenamefont{Biswas et~al.}(2000)\citenamefont{Biswas, Cain, Roemer,
  and Schreiber}}]{Biswas}
\bibinfo{author}{\bibfnamefont{P.}~\bibnamefont{Biswas}},
  \bibinfo{author}{\bibfnamefont{P.}~\bibnamefont{Cain}},
  \bibinfo{author}{\bibfnamefont{R.~A.} \bibnamefont{Roemer}},
  \bibnamefont{and}
  \bibinfo{author}{\bibfnamefont{M.}~\bibnamefont{Schreiber}},
  \bibinfo{journal}{Phys. Status Solidi B} \textbf{\bibinfo{volume}{218}},
  \bibinfo{pages}{205} (\bibinfo{year}{2000}).

\bibitem[{\citenamefont{Fleishman and Licciardello}(1977)}]{Fleishman}
\bibinfo{author}{\bibfnamefont{L.}~\bibnamefont{Fleishman}} \bibnamefont{and}
  \bibinfo{author}{\bibfnamefont{D.~C.} \bibnamefont{Licciardello}},
  \bibinfo{journal}{J. Phys. C} \textbf{\bibinfo{volume}{10}},
  \bibinfo{pages}{L125} (\bibinfo{year}{1977}).

\bibitem[{\citenamefont{Soukoulis and Economou}(1981)}]{Economou}
\bibinfo{author}{\bibfnamefont{C.~M.} \bibnamefont{Soukoulis}}
  \bibnamefont{and} \bibinfo{author}{\bibfnamefont{E.~N.}
  \bibnamefont{Economou}}, \bibinfo{journal}{Phys. Rev. B}
  \textbf{\bibinfo{volume}{24}}, \bibinfo{pages}{5698} (\bibinfo{year}{1981}).

\bibitem[{\citenamefont{Kozlov et~al.}(1998)\citenamefont{Kozlov, Malyshev,
  Dom\'{\i}nguez-Adame, and Rodr\'{\i}guez}}]{Kozlov}
\bibinfo{author}{\bibfnamefont{G.~G.} \bibnamefont{Kozlov}},
  \bibinfo{author}{\bibfnamefont{V.~A.} \bibnamefont{Malyshev}},
  \bibinfo{author}{\bibfnamefont{F.}~\bibnamefont{Dom\'{\i}nguez-Adame}},
  \bibnamefont{and}
  \bibinfo{author}{\bibfnamefont{A.}~\bibnamefont{Rodr\'{\i}guez}},
  \bibinfo{journal}{Phys. Rev. B} \textbf{\bibinfo{volume}{58}},
  \bibinfo{pages}{5367} (\bibinfo{year}{1998}).

\bibitem[{\citenamefont{Sloane}()}]{Wythoff}
\bibinfo{author}{\bibfnamefont{N.~J.~A.} \bibnamefont{Sloane}},
  \emph{\bibinfo{title}{The on-line encyclopedia of integer sequences}},
  \bibinfo{howpublished}{\url{https://oeis.org/A000201}},
  \bibinfo{note}{accessed: 2016-06-27}.

\end{thebibliography}
    \end{document}